\documentclass[final,3p,times]{elsarticle}

\usepackage{lineno}

\usepackage{multibib}
\usepackage{natbib}
\usepackage{graphicx}
\usepackage{tabularx}
\usepackage{paralist}
\usepackage{soul}
\usepackage{amsmath}
\usepackage{amsfonts}
\usepackage{amssymb}
\usepackage{amsthm}
\usepackage{multirow}
\usepackage{subcaption}
\usepackage{ulem} 
\usepackage[utf8]{inputenc}
\usepackage{hyperref}
\usepackage{xstring}
\usepackage{xcolor}
\usepackage[inline]{enumitem}
\usepackage[skip=1ex]{caption}

\journal{Neurocomputing}

\begin{document}

\makeatletter
\def\ps@pprintTitle{%
  \let\@oddhead\@empty
  \let\@evenhead\@empty
  \def\@oddfoot{\reset@font\hfil\thepage\hfil}
  \let\@evenfoot\@oddfoot
}
\makeatother

\begin{frontmatter}



\title{Whose Emotion Matters? Speaking Activity Localisation without Prior Knowledge}


\author[1]{Hugo Carneiro}
\ead{hugo.carneiro@uni-hamburg.de}
\author[1]{Cornelius Weber}
\ead{cornelius.weber@uni-hamburg.de}
\author[1]{Stefan Wermter}
\ead{stefan.wermter@uni-hamburg.de}

\affiliation[1]{organization={University of Hamburg, Department of Informatics},
            addressline={Vogt-Koelln-Str.~30},
            city={Hamburg},
            postcode={22527},
            country={Germany}}

\begin{abstract}
    The task of emotion recognition in conversations (ERC) benefits from the availability of multiple modalities, as provided, for example, in the video-based \textbf{M}ultimodal \textbf{E}motion\textbf{L}ines \textbf{D}ataset (MELD). However, only a few research approaches use both acoustic and visual information from the MELD videos. There are two reasons for this: First, label-to-video alignments in MELD are noisy, making those videos an unreliable source of emotional speech data. Second, conversations can involve several people in the same scene, which requires the localisation of the utterance source. In this paper, we introduce \textbf{MELD} with \textbf{F}ixed \textbf{A}udiovisual \textbf{I}nformation via \textbf{R}ealignment (MELD-FAIR) by using recent active speaker detection and automatic speech recognition models, we are able to realign the videos of MELD and capture the facial expressions from speakers in 96.92\% of the utterances provided in MELD. Experiments with a self-supervised voice recognition model indicate that the realigned MELD-FAIR videos more closely match the transcribed utterances given in the MELD dataset. Finally, we devise a model for emotion recognition in conversations trained on the realigned MELD-FAIR videos, which outperforms state-of-the-art models for ERC based on vision alone. This indicates that localising the source of speaking activities is indeed effective for extracting facial expressions from the uttering speakers and that faces provide more informative visual cues than the visual features state-of-the-art models have been using so far. The MELD-FAIR realignment data, and the code of the realignment procedure and of the emotional recognition, are available at \url{https://github.com/knowledgetechnologyuhh/MELD-FAIR}. 
\end{abstract}




\end{frontmatter}



    \section{Introduction}
    \label{sec:intro}
    
        Emotion recognition in conversations (ERC) is a task that involves recognising the emotion of interlocutors in a dialogue. Challenges of this task include the modelling of the conversational context and how the emotion of the interlocutors may change depending on that context, which is called emotion shift~\citep{poria2019}. ERC can prove helpful in real-world scenarios in which people are talking with each other, for example, in human-robot interaction applications~\citep{hornecker2022,krummheuer2020,utami2019}. However, most ERC datasets are exclusively based on text transcriptions of conversations~\citep{hsu2018,li2017,zahiri2018} or are restricted to dyadic interactions in very controlled environments~\citep{busso2008,gary2012}.
        
        \citet{poria2019} published the first large-scale multimodal ERC dataset with several interlocutors, the \textbf{M}ultimodal \textbf{E}motion\textbf{L}ines \textbf{D}ataset (MELD). The dataset consists of videos extracted from the \textit{Friends} TV series. Each video is cut to match a single utterance, and the videos are organised into dialogues and utterances, with each dialogue having one or more utterances. Together with the acoustic and visual information provided by the videos, the text transcription of every utterance and the speaker label are also provided.
        
\begin{figure}[!ht]
            \centering
            \subfloat[Original video cuts falsely corresponding to two consecutive utterances. The expected corresponding utterances are given below each video.\label{fig:unaligned}]{
                \includegraphics[width=0.88\textwidth]{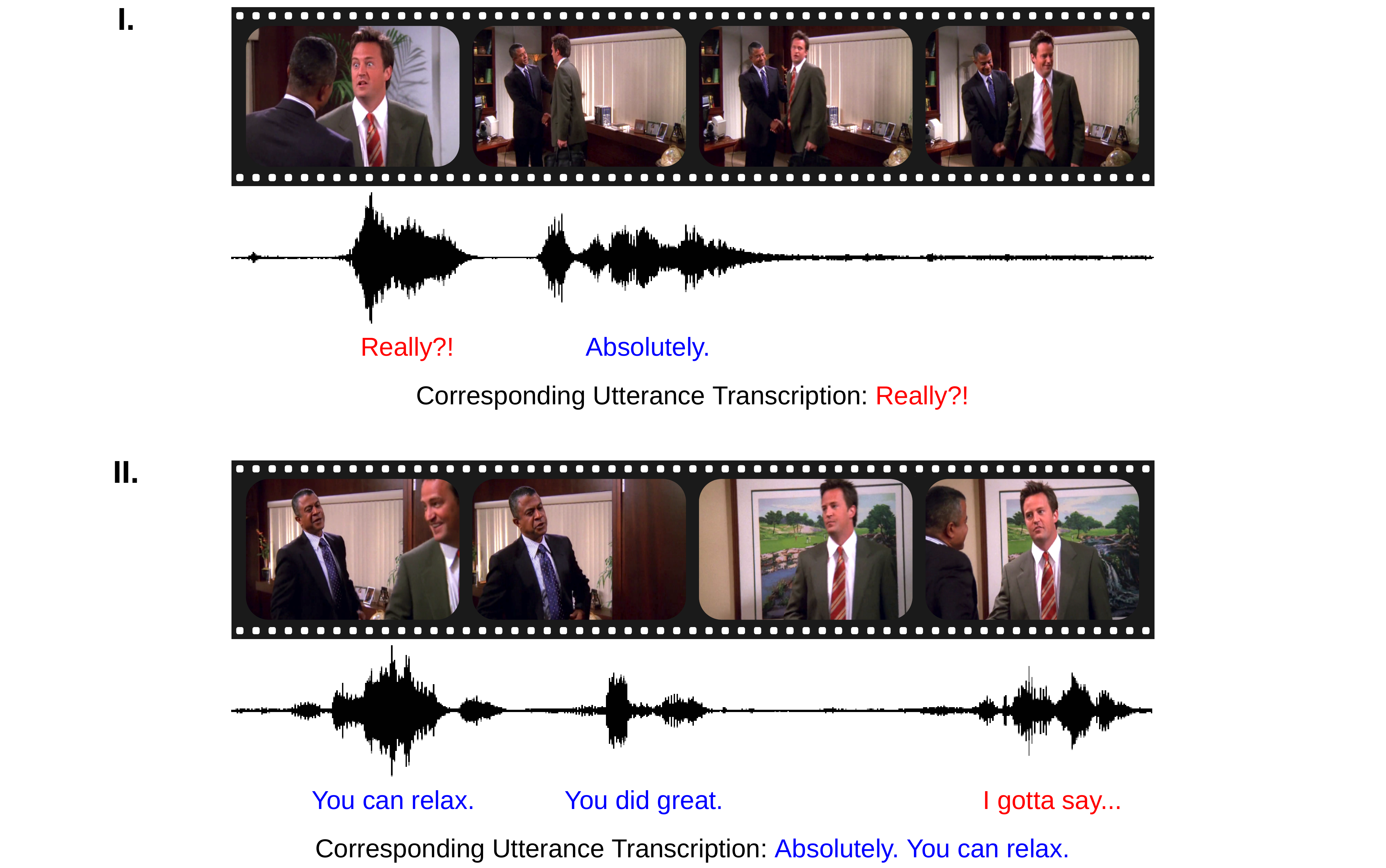}
            }
            
            \bigskip
            
            \subfloat[Realigned video cuts, correctly matching the corresponding utterances.\label{fig:aligned}]{
                    \includegraphics[width=0.88\textwidth]{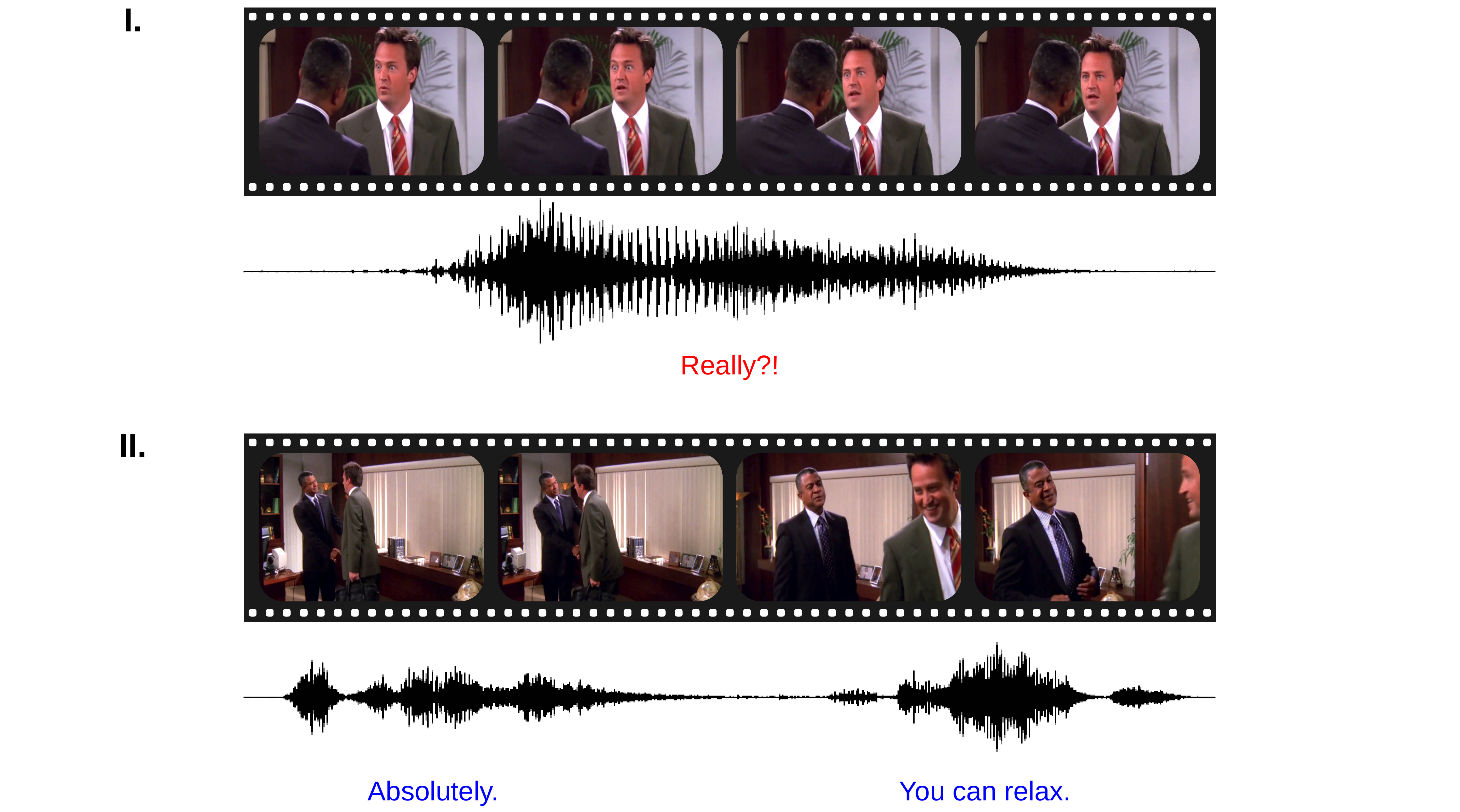}
            }
            \caption{Example of misaligned video cuts provided in the MELD dataset, and their corresponding correction. The different colours in the utterances represent the different speakers in the video cuts.}
            \label{fig:meld_misalignment_example}
        \end{figure}

        Many approaches have been proposed to tackle the task of ERC in MELD. Even though MELD was created to be a multimodal dataset, most of the approaches rely exclusively on textual information~\citep{ghosal2020,lee2021,lee2022,saxena2022,song2022,zhu2021}. Using the visual modality is difficult due to
        frequent misalignments between video cuts and the expected corresponding utterances (see Figure~\ref{fig:meld_misalignment_example} for an example). This is likely a consequence of an automatic generation of the video cuts with the \textit{Gentle}\footnote{\url{https://lowerquality.com/gentle/}} transcription alignment tool.
        
        For some years, there has been a demand for more reliable information from the visual modality, given the frequent problems of video-text synchronisation\footnote{\url{https://github.com/declare-lab/MELD/issues/9}}. Video cuts and utterance transcriptions can be misaligned in a variety of ways. Figure~\ref{fig:unaligned} presents two cases of misalignment. In case I, the utterance appears within the first half of the video cut and another person's utterance is falsely assigned to the same cut. In case II, the utterance starts being spoken in the video cut assigned to the preceding utterance and continues through the first half of the video cut assigned to that target utterance. Figure~\ref{fig:aligned} depicts the corrected alignment between the video cuts and their corresponding utterance transcriptions. This is a result of our dataset refinement procedure (cf. Section~\ref{sec:procedure_description}).
        
        Facial expressions and speech signals provide relevant information regarding the emotion of a person. However, the noticeable number of mismatching cases between video cuts and the corresponding utterance transcriptions hindered the use of those modalities for some years, with information from the visual modality being disregarded even by the dataset creators, who stated that video-based speaker localisation were still open problems~\citep{poria2019}. Accordingly, in the dataset itself, no information on the location of the face of the uttering speakers is offered.
        
        Even though quite rarely, speech data from the videos of MELD has been used for ERC since the work of \citet{poria2019}. However, without the proper alignment correction of the videos, audio samples used for this task can include speech from other speakers with different emotions. In contrast, only quite recently there has been some interest in the use of visual information from the MELD videos~\citep{chudasama2022,hu2022,hu2021,li2022,xie2021}. However, alongside the problems that arise with the lack of proper realignments, the proposed solutions do not take into account the necessity to localise the source of the speaking activity in a particular scene or frame, which, in turn, is useful to extract of the emotional facial expressions of the uttering speaker. The added information from acoustic and visual modalities has improved ERC compared to models that use information obtained exclusively from utterance transcriptions. However, those improvements are limited because of the unreliability of those modalities.
        
        Recent advances in active speaker detection (ASD) in the wild~\citep{alcazar2020,alcazar2021,carneiro2021,min2022,tao2021,zhang2021} indicate the capability of audiovisual neural models to localise sources of speaking activity in videos given the faces of the people as well as the audio of a scene. Localising the active speaker can enable more reliable emotion recognition from video in MELD. State-of-the-art ASD models can be very precise in determining who among multiple people is speaking, especially if there are at least a few seconds of continuous speaking activity. Multi-party scenarios can still present challenges in accurately localising the source of some particular speaking activity. These challenges include:
        \begin{inparaenum}[\itshape i\upshape )]
            \item the partial occlusion of the speaker's face by objects or other people;
            \item the presence of other people in the same scene moving their mouths, even though they are not actively speaking;
            \item interfering noise, such as background chatting or, in the case of TV sitcoms, laugh tracks; and
            \item the active speaker not being in the main focus of the scene, and having the speaker's face in a considerably smaller size and resolution than other non-speaking people.
        \end{inparaenum}
        
\begin{figure}[!ht]
            \centering
            \hspace*{\fill}
            \subfloat[Three people interacting, with two of them having their faces visible and only one speaking.\label{fig:asd_ava_activespeaker_few_people}]{
                \includegraphics[height=0.165\textheight]{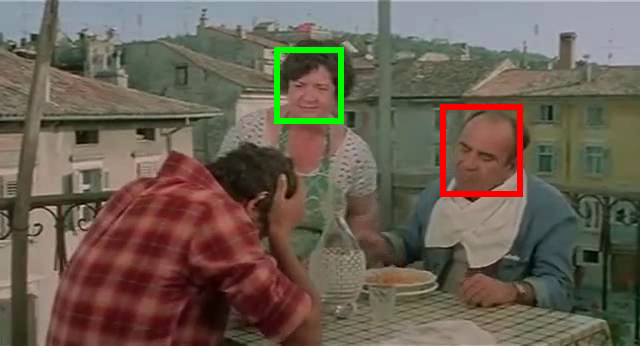}
            }\hfill
            \subfloat[Scene with several people, with two of them speaking simultaneously.\label{fig:asd_ava_activespeaker_many_people}]{
                \includegraphics[height=0.165\textheight]{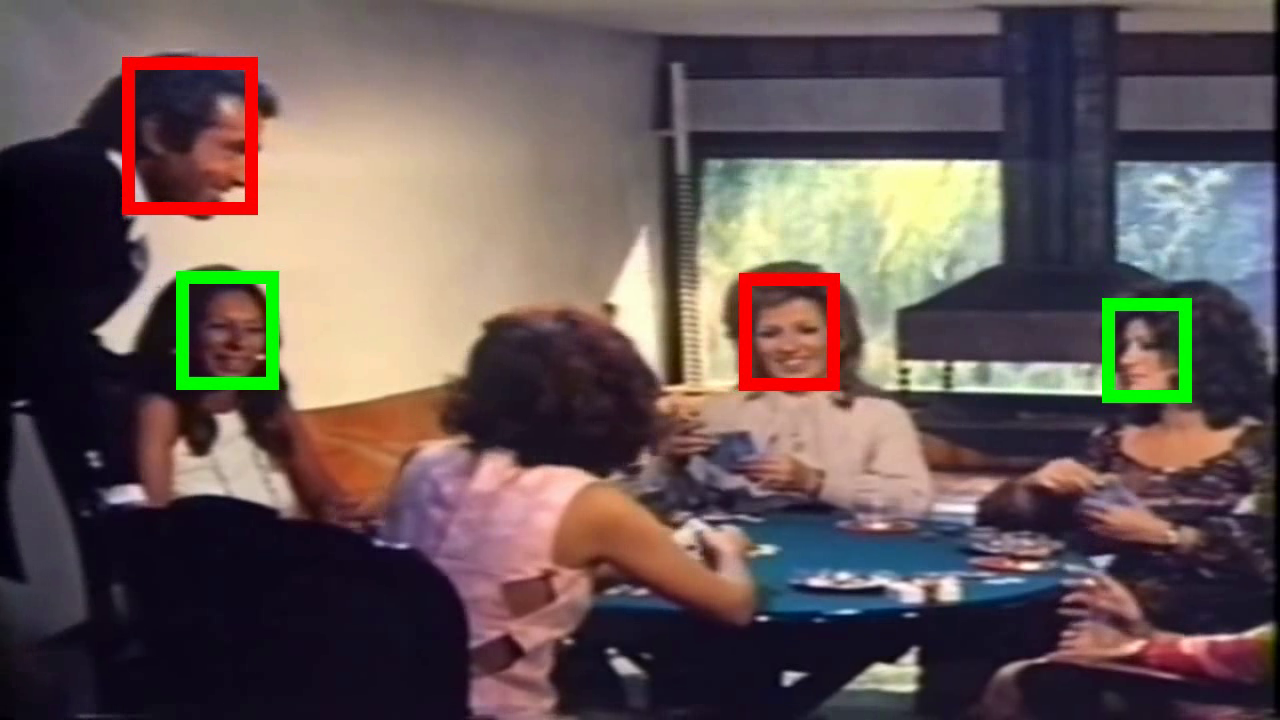}
            }
            \hspace*{\fill}
            \caption{Examples of conversational scenes from AVA-ActiveSpeaker videos. Green boxes identify those who are speaking, whereas red boxes mark silent people.}
            \label{fig:asd_ava_activespeaker_examples}
        \end{figure}

        Most in-the-wild ASD models were trained on AVA-ActiveSpeaker, a dataset containing videos in a large variety of resolutions~\citep{roth2020}. The videos of AVA-ActiveSpeaker contain scenes with multiple people speaking with each other, which is similar to the conversational scenes in the videos of MELD. Figure~\ref{fig:asd_ava_activespeaker_examples} displays examples of conversational scenes present in the videos of the AVA-ActiveSpeaker dataset.
        
        The first contribution of this paper is to offer a new method to extract the position of the faces of active speakers for datasets, which can be useful for tasks in which the facial information may provide additional relevant information, but, for some reason, the face position is not given in the dataset. The procedure can be used in any dataset with humans speaking without annotation concerning the visual modality, e.g., the position of the speaker's face. The second contribution of this paper is the evaluation of this procedure on the MELD dataset, and the consequent development of a refined version of MELD, named MELD with Fixed Audiovisual Information via Realignment (MELD-FAIR). Finally, to assess the applicability of the extraction of the faces of active speakers for the task of ERC, we propose an emotion recognition model whose outstanding performance on the visual data indicates that the faces extracted from the active speakers indeed provide an informative visual cue for the task of ERC.
        
        The paper is structured as follows. Section~\ref{sec:meld} offers a brief overview and some specific details on the MELD dataset. Section~\ref{sec:procedure_description} describes the procedure of dataset refinement, which consists of correcting the alignment between video cuts and the corresponding utterances, and determining the position of the face of the uttering speaker in each frame of the newly produced video cut. Section~\ref{sec:dataset_assessment} provides quantitative analysis of the resulting dataset, comparing it with the characteristics of the original dataset that were provided in Section~\ref{sec:meld}. In that same section, experiments are also provided, as a means to evaluate how well the resulting dataset applies to the task of emotion recognition. Section~\ref{sec:discussion} discusses the results.
    
    \section{The MELD Dataset}
    \label{sec:meld}
    
        MELD contains scenes from various episodes of the \textit{Friends} TV series. Those scenes are denoted as dialogues, and each dialogue is organised as a sequence of utterances. For every utterance, there is a corresponding dataset entry containing the speaker's identity, emotion and sentiment. The annotated emotion can be either one of Ekman's universal emotions (\textit{joy}, \textit{sadness}, \textit{fear}, \textit{anger}, \textit{disgust}, and \textit{surprise}), or \textit{neutral} if no particular emotion was noticed by the dataset annotators.
        
        MELD is split into three sets, denoted \textit{train}, \textit{dev}, and \textit{test}. Each data record in those splits contains the following information: the utterance, its speaker, the emotion perceived in that utterance, the corresponding sentiment, a dialogue identifier, an utterance identifier, the season and episode of \textit{Friends} in which that scene happened, a time stamp determining where that scene starts, as well as one determining where that scene ends. For every split, a dataset record can be uniquely identified by its dialogue identifier and its utterance identifier.
        
        Table~\ref{tab:meld_dataset_conv_excerpt} presents an excerpt of a conversation, containing a sequence of contiguous data records and corresponding labels for the uttering speaker, his or her emotion, and the corresponding sentiment. The misalignment of the video cuts can provide overlaps, which are indicated by the start and end time stamps. Table~\ref{tab:meld_dataset_conv_excerpt_overlap_start_end_times} indicates that two videos of consecutive utterances present an overlap due to a wrongly executed alignment process.
        
        \begin{table}[ht!]
            \centering
            \caption{Excerpt of a dyadic conversation from MELD \textit{train} split with corresponding speaker, emotion, and sentiment information}
            \label{tab:meld_dataset_conv_excerpt}
            \begin{tabular}{cc|p{4.25cm}ccc}
                Dia & Utt & \centering Utterance & Speaker & Emotion & Sentiment \\ \hline
                D0 & U5 & Now you'll be heading a whole division, so you'll have a lot of duties. & Interviewer & neutral & neutral \\
                D0 & U6 & I see. & Chandler & neutral & neutral \\
                D0 & U7 & But there'll be perhaps 30 people under you, so you can dump a certain amount on them. & Interviewer & neutral & neutral \\
                D0 & U8 & Good to know. & Chandler & neutral & neutral \\
                D0 & U9 & We can go into detail. & Interviewer & neutral & neutral \\
                D0 & U10 & No, don't. I beg of you! & Chandler & fear & negative
            \end{tabular}
        \end{table}
        
        \begin{table}[ht!]
            \centering
            \caption{Additional information in MELD about the utterances presented in Table~\ref{tab:meld_dataset_conv_excerpt}. Overlaps between video cuts due to mistaken determination of the start and end times of an utterance are marked in \textbf{bold}}
            \label{tab:meld_dataset_conv_excerpt_overlap_start_end_times}
            \begin{tabular}{cc|cccc}
                Dia & Utt & Season & Episode & Start time & End time\\ \hline
                D0 & U5 & S8 & E21 & 0:16:41.126 & 0:16:44.337 \\
                D0 & U6 & S8 & E21 & 0:16:48.800 & \textbf{0:16:51.886} \\
                D0 & U7 & S8 & E21 & \textbf{0:16:48.800} & 0:16:54.514 \\
                D0 & U8 & S8 & E21 & 0:16:59.477 & 0:17:00.478 \\
                D0 & U9 & S8 & E21 & 0:17:00.478 & 0:17:02.719 \\
                D0 & U10 & S8 & E21 & 0:17:02.856 & 0:17:04.858
            \end{tabular}
        \end{table}

    \section{Dataset Refinement Procedure}
    \label{sec:procedure_description}
        
\begin{figure}[ht!]
            \centering
            \subfloat[Realignment of MELD videos.\label{fig:video_realignment}]{
            		   \includegraphics[width=0.72\textwidth]{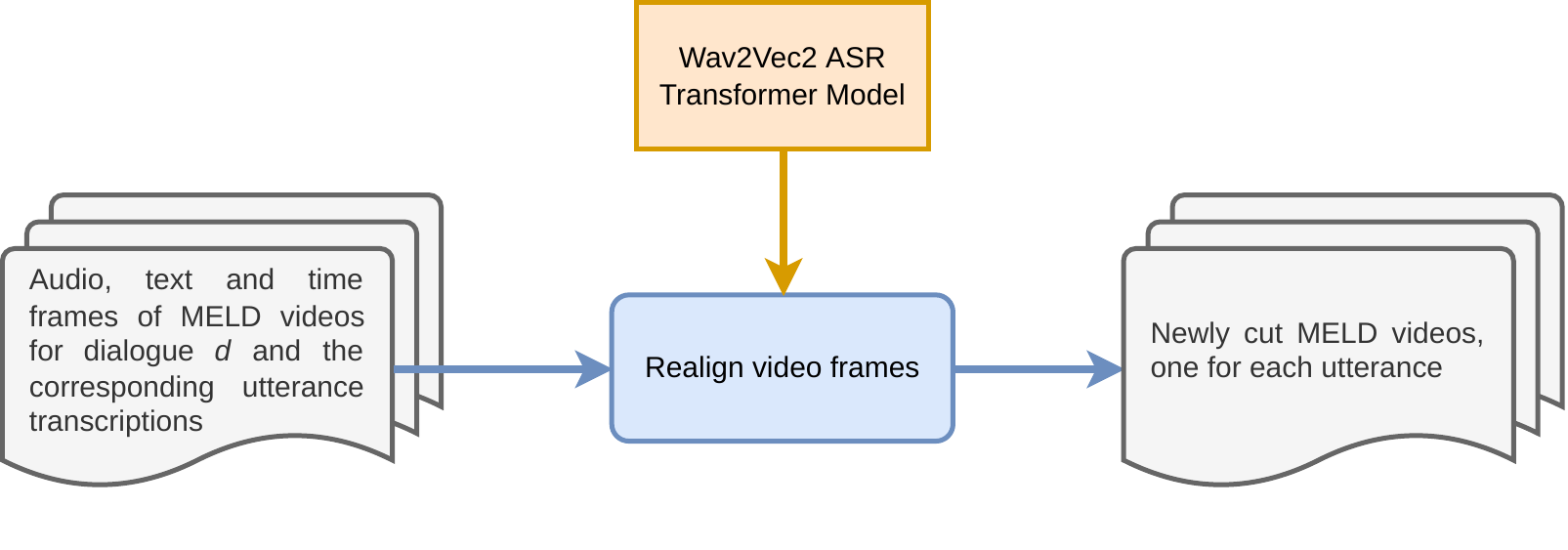}
            }
            
            \bigskip
            
            \subfloat[Extraction of visual and acoustic data.\label{fig:extract_faces_and_audio}]{
            		   \includegraphics[width=0.9\textwidth]{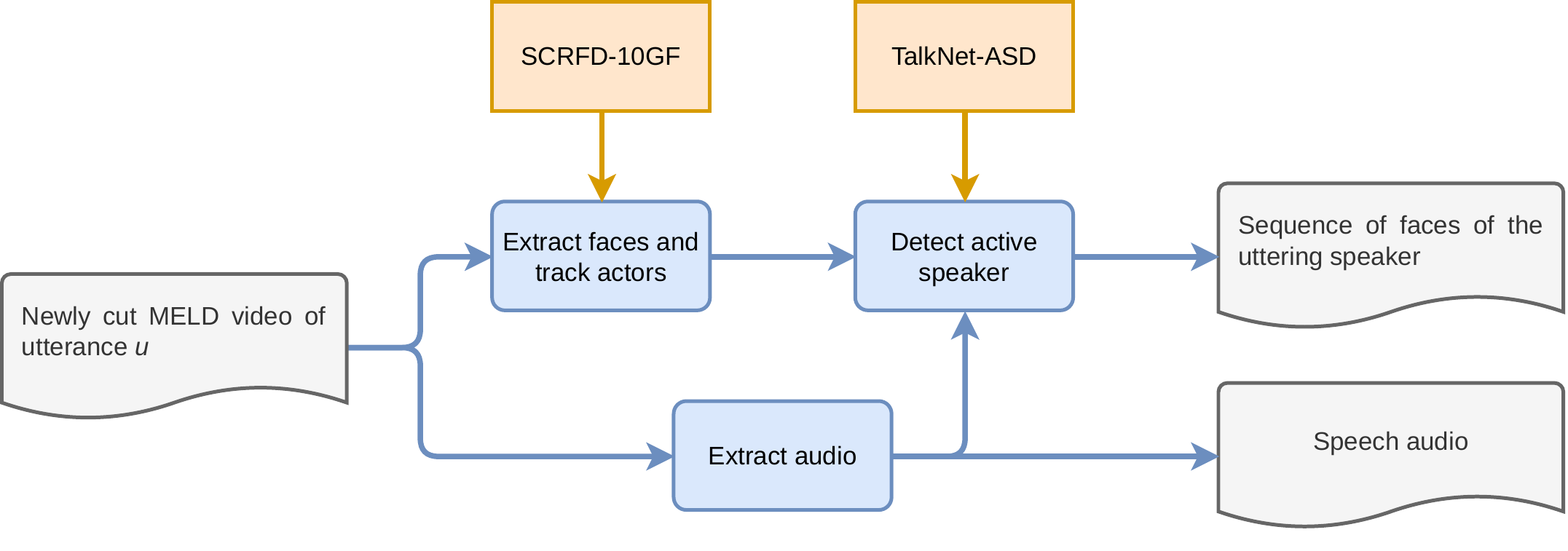}
            }
            \caption{Major steps of the dataset refinement procedure. Orange arrows indicate the application of a model, and blue arrows represent information flow.}
            \label{fig:dataset_refinement}
        \end{figure}

        The extraction of emotional speech and emotional facial expressions depends on having audio samples that match closely enough the utterance being said and on being capable of localising the uttering speaker in a scene, particularly that person's face. To meet both requirements, the dataset refinement procedure is divided into two parts, with each part addressing one requirement. First, the videos of MELD are realigned, such that their audios match closely enough the target utterance, as indicated in the flow chart in Figure~\ref{fig:video_realignment}. Next, with the videos properly realigned, the faces of the people in the scene are extracted and organised into sequences. Then, given the extracted sequences of faces and the scene audio, an ASD model determines which of these sequences corresponds to the uttering speaker (see the flow chart in Figure~\ref{fig:extract_faces_and_audio}). The resulting set including the realigned audio from all videos and the sequence of facial expressions of the person vocalising in each of those videos constitutes a refined version of MELD called MELD-FAIR.
        
        \subsection{Video Realignment}
        \label{sec:video_realigment}
        
            Each video in the MELD dataset corresponds to a particular utterance, which, in turn, belongs to a sequence of utterances, also called a dialogue. Videos that are misaligned to their corresponding utterances are a consequence of the mistaken determination of where the boundaries of those particular utterances lie within their respective dialogues. A considerable number of misaligned videos may prevent the proper identification of the source of speaking activity, especially because sometimes the speaking activity might happen partially in the target video and partially in the one that precedes or in the one that follows it in the dialogue (see the example depicted in Figure~\ref{fig:meld_misalignment_example}).
            
            The realignment of the videos takes into account that videos that belong to the same dialogue are organised sequentially. First, the audio signal of every dataset video is extracted. Next, for every split $\sigma \in \left\{ \text{\textit{train}}, \text{\textit{dev}}, \text{\textit{test}} \right\}$ and dialogue $d$, the audio signals $a_{\sigma, d, u}$ corresponding to each utterance $u$ belonging to dialogue $d$ are concatenated in order. Existing overlaps, such as the one indicated in Table~\ref{tab:meld_dataset_conv_excerpt_overlap_start_end_times}, are removed by truncating the audio signals that lead to those overlaps. Silence blocks are added between consecutive video cuts if there is a time difference between the end time stamp of a video and the start time stamp of the following. The length of a silence block is equal to the corresponding time difference, but long silence blocks are capped at 250 ms. Due to a few videos whose length is much longer than their corresponding utterances, video lengths are also capped at 45 seconds. This affects two of altogether 13,708 videos (check \ref{sec:app_problematic} for an indication of the videos affected by the 45-second capping).
            
\begin{figure}[ht!]
                \centering
                \includegraphics[width=0.9\textwidth]{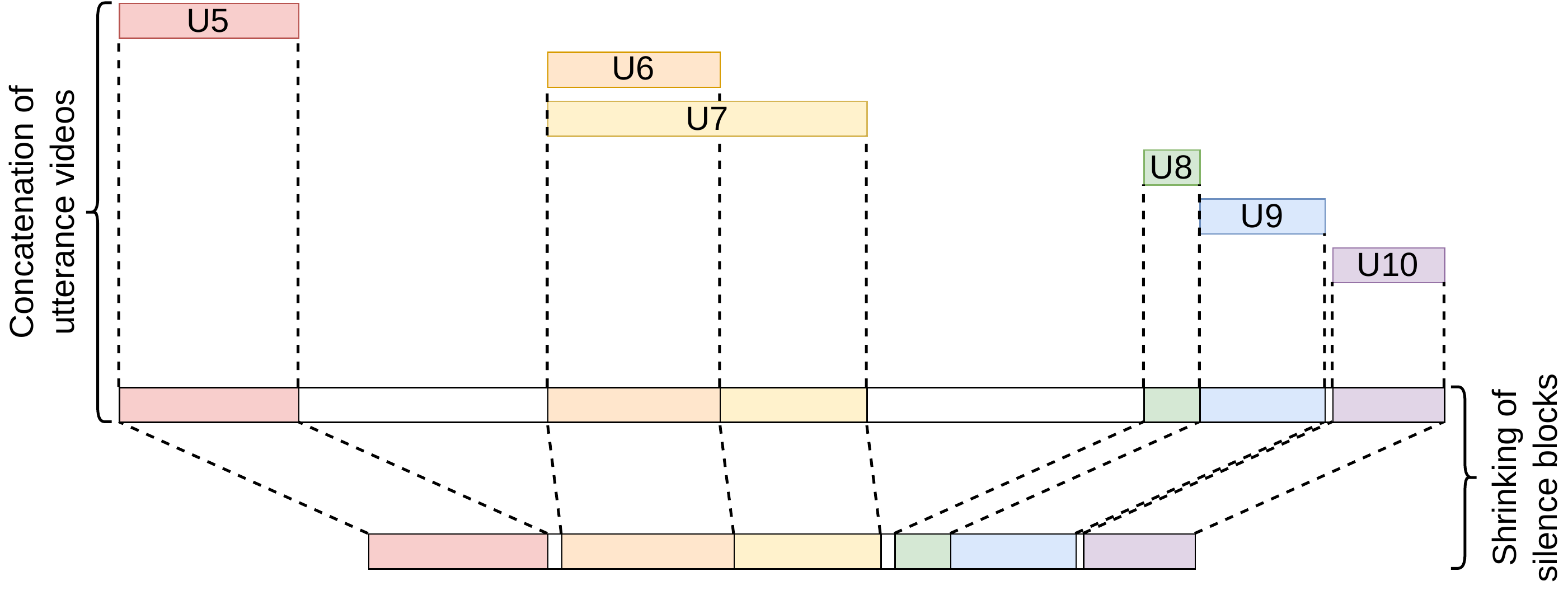}
                \caption{Schematic representation of the concatenation of the audio signals of a dialogue. First, the audios of all utterances of a given dialogue are concatenated, with silence blocks inserted wherever it is adequate. Next, the lengths of the silence blocks are reduced to a minimum length that still allows for the identification of individual blocks of consecutive utterances (e.g., utterances U6 and U7, and U8 and U9).}
                \label{fig:audio_signal_concatenation}
            \end{figure}

            Figure~\ref{fig:audio_signal_concatenation} presents a graphical representation of the concatenation of audio signals. Each box labelled U5 to U10 represents the audio signal of an utterance. The lengths of those boxes are proportional to the duration of the utterances presented in Tables~\ref{tab:meld_dataset_conv_excerpt} and~\ref{tab:meld_dataset_conv_excerpt_overlap_start_end_times}, which can be inferred by their start and end time stamps. The label used in each box corresponds to the utterance identifier given in Table~\ref{tab:meld_dataset_conv_excerpt_overlap_start_end_times}. The gaps between the boxes are proportional to the distance between the end of an utterance and the beginning of the following one. The figure presents an example of overlap being removed by altering the start time of utterance U7. It also shows the insertion of silence blocks where the gaps between utterances lie, and the subsequent capping of silence block lengths to 250 ms.
            
            The utterance transcriptions are concatenated as well. Prior to their concatenation, all punctuation marks in each transcription are removed, and both a start-of-sequence and an end-of-sequence token are appended to each end of every utterance transcription within a dialogue. With the audio signals and the transcriptions properly concatenated, the text of the concatenated transcription is aligned to the concatenated audio through forced alignment using connectionist-temporal-classification (CTC) segmentation~\cite{kuerzinger2020}. Given a speech audio signal, CTC segmentation uses frame-based character posterior probabilities generated by a CTC-based end-to-end network. From these character-level probabilities, maximum joint probabilities are computed via dynamic programming. These maximum joint probabilities indicate how likely a given excerpt from the dialogue transcription is aligned to a particular slice of the speech audio signal. After the maximum joint probability for the alignment of the complete dialogue transcription to the whole speech audio signal is computed, the character-wise alignment is obtained by backtracking from the most probable temporal position of the last character in the transcription. The CTC-based end-to-end network used to generate the character-level probabilities had to be pretrained on already aligned data, for which the Wav2Vec2~\citep{baevski2020} automatic speech recognition transformer model\footnote{More specifically, the Wav2Vec2 Large (LV-60) model pretrained and fine-tuned on 960 hours of speech audio from Libri-Light and Librispeech (see list of pretrained models at \url{https://github.com/facebookresearch/fairseq/tree/main/examples/wav2vec}). }~\citep{vaswani2017} was used.
            
            The video realignment procedure is executed for each dialogue in the dataset. Most of the processing time is dedicated to the generation of frame-based character posterior probabilities by the CTC-based end-to-end network and the subsequent computation of maximum joint probabilities. The former is run on graphical processing units (GPUs) with high parallelisation capabilities. The latter involves a dynamic programming algorithm whose processing time is proportional to the number of audio frames of the whole dialogue and to the square of the number of characters of the concatenated utterance transcriptions.

        \subsection{Uttering Speaker Localisation}
        \label{sec:uttering_speaker_localisation}
            
            With videos that very likely contain the part of a scene in which a given utterance is said, it is possible to localise the source of the speaking activity, i.e., the person who spoke the utterance. Figure~\ref{fig:extract_faces_and_audio} schematically represents the process of extracting the speech audio as well as face images of the uttering speaker from a video. As a first step, an efficient face detection model with sample and computation redistribution (SCRFD-10GF)~\citep{guo2022} is used to detect all faces in every frame of those videos. Faces detected this way are then subsequently extracted and organised into ordered groups, creating several sequences of faces. Each face is identified by the video frame from which it is extracted, and by an identifier of the sequence it belongs to. For the organisation of the faces into sequences, faces detected in consecutive frames are considered to belong to the same sequence if the intersection-over-union (IoU) ratio between their areas is greater than a given threshold $\theta$. In case there is more than one pair of faces extracted from consecutive frames that satisfy this condition, the face pair with the highest IoU ratio is considered as belonging to the same sequence.
            
\begin{figure}[!ht]
                \centering
                \includegraphics[width=0.9\textwidth]{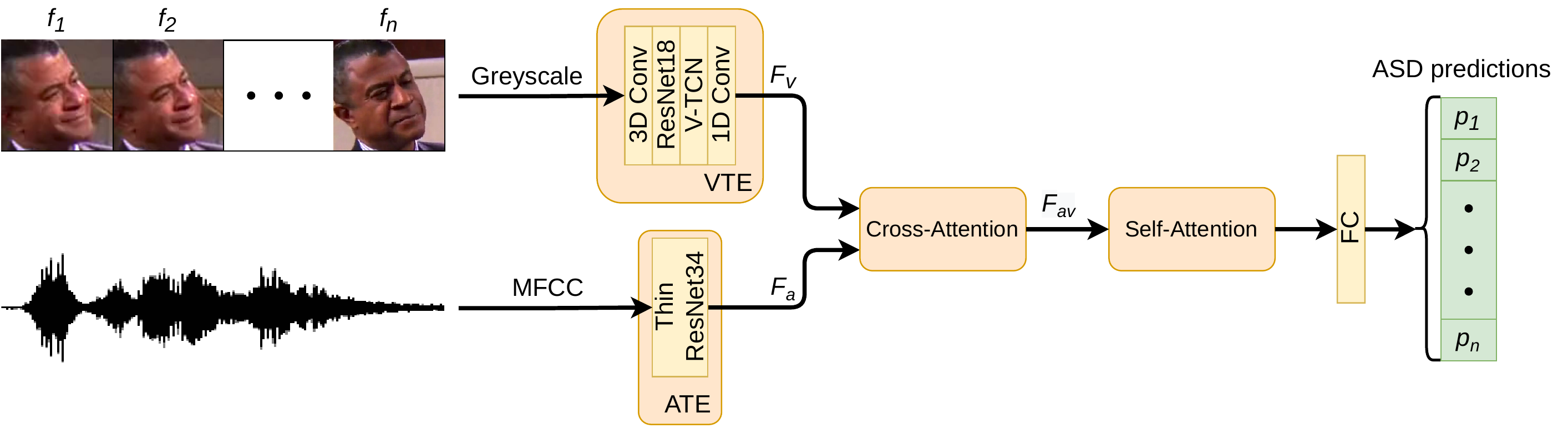}
                \caption{TalkNet-ASD architecture}
                \label{fig:TalkNet-ASD_architecture}
            \end{figure}

            Each face sequence and the corresponding slice of the speech signal is then sent to TalkNet-ASD~\citep{tao2021}, an audiovisual ASD model, to determine whether that face sequence presents some indication of speaking activity that resembles that slice of the speech signal. Figure~\ref{fig:TalkNet-ASD_architecture} shows a sketch of TalkNet-ASD's architecture. TalkNet-ASD uses a visual temporal encoder (VTE) to learn long-term representations of facial expression dynamics, and an audio temporal encoder (ATE) to learn audio content representations from the temporal dynamics~\citep{tao2021}.
            
            VTE consists of a front end, where video frame streams are encoded into sequences of frame-based embeddings, and a visual temporal network, whose aim is to represent the temporal content in a long-term spatiotemporal structure~\citep{tao2021}. Its front end is based on the vision module introduced in~\citep{afouras2018}, consisting of a 3D convolution layer with a filter width of 5 frames followed by a 2D 18-layer residual network. Given an input with dimensions $T_{v} \times C \times W \times H$, where $T_{v}$ is the number of frames, and $C$, $W$ and $H$ are the number of channels, width and height of each frame, the front end yields a tensor with dimensions $T_{v} \times \frac{W}{32} \times \frac{H}{32} \times 512$, which is subsequently average-pooled in both its spatial dimensions, thus producing a feature vector with 512 dimensions for each input frame. Similarly to the visual model of \citet{afouras2018}, TalkNet-ASD receives a sequence of greyscale images, which means that the number of channels $C$ in each frame is 1. TalkNet-ASD's visual temporal network (V-TCN) consists of a 5-block residual network followed by a sequence of two 1D convolution layers. The residual blocks consist of a 1D depth-separable convolution layer followed by rectifier linear units and batch normalisation layers. The residual network is responsible for obtaining a representation of the temporal content. The representation consists of a tensor with dimensions $T_{v} \times 512$. The sequence of 1D convolution layers finally reduces the dimensionality of this tensor, yielding a visual embedding $F_{v}$ of dimensions $T_{v} \times 128$, i.e., 128 dimensions for every input frame.
            
            The speech signal is first encoded as a sequence of overlapping audio frames, each one characterised by a 13-dimensional vector of Mel-frequency cepstral coefficients (MFCCs) based on a window size of 25 ms and a window step of 10 ms. This means that given a sequence of $T_{a}$ audio frames, ATE receives as input a tensor with dimensions $1 \times 13 \times T_{a}$. ATE consists of a 2D 34-layer residual network with squeeze-and-excitation (SE) modules~\citep{hu2018}. The number of channels in each block of the ResNet34 network is also reduced to one quarter of the number in each block of the original ResNet with 34 layers, similarly to the Thin ResNet34 introduced by \citet{chung2020}. The output of the audio encoder is an audio embedding $F_{a}$ of dimensions $\frac{T_{a}}{4} \times 128$. The dimensions of $F_{a}$ and $F_{v}$, the embeddings output by both encoders, match when the number of audio frames is equal to four times the number of visual frames (or face crops). The matching in their dimensions is a necessary feature for the subsequent attention mechanism. A direct implication of the number of audio frames being four times the number of video frames is that each video frame corresponds to roughly 40 milliseconds of the video (or 25 fps) since the length of the window step between consecutive overlapping audio frames is 10 milliseconds.
            
            With the motivation of audiovisual synchronisation working as an informative cue for speaking activities, TalkNet-ASD contains a cross-attention subnetwork that receives $F_{a}$ and $F_{v}$ as inputs, and outputs an audio attention feature $F_{a \rightarrow v}$ and a video attention feature $F_{v \rightarrow a}$. $F_{a \rightarrow v}$ is obtained through the application of $F_{v}$ as the target sequence to generate the query $Q_{v}$ in the attention layer and $F_{a}$ as the source sequence to generate key $K_{a}$ and value $V_{a}$. $F_{v \rightarrow a}$ is obtained through an analogous process. Next, $F_{a \rightarrow v}$ and $F_{v \rightarrow a}$ are concatenated into a single audiovisual attention feature vector $F_{av}$ which is sent to a self-attention subnetwork whose aim is to model audiovisual utterance-level information, and this way distinguish between speaking and non-speaking frames. Both cross-attention and self-attention subnetworks contain one transformer layer with eight attention heads each~\citep{vaswani2017}.
            
\begin{figure}[!ht]
                \centering
                \hspace*{\fill}
                \subfloat[Conversation between three people in a low-lighting environment.\label{fig:asd_meld_few_people}]{
                    \includegraphics[height=0.165\textheight]{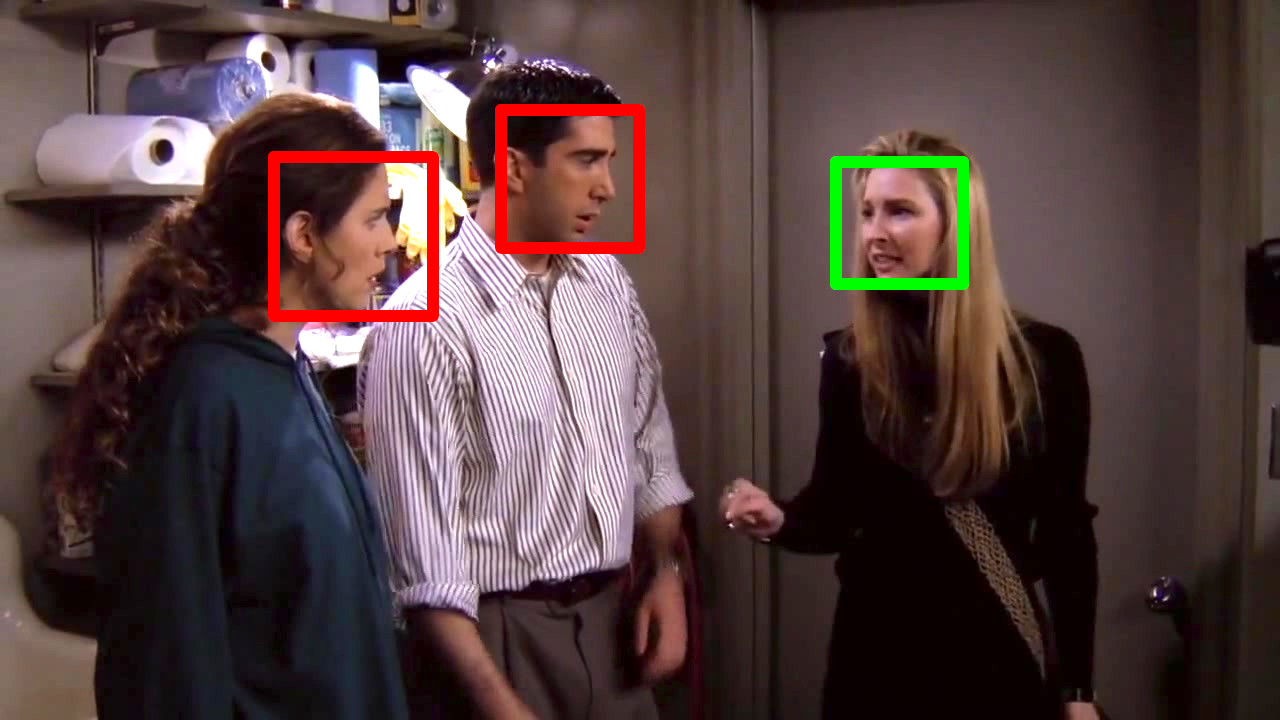}
                }\hfill
                \subfloat[Conversational scene in a crowded environment.\label{fig:asd_meld_many_people}]{
                        \includegraphics[height=0.165\textheight]{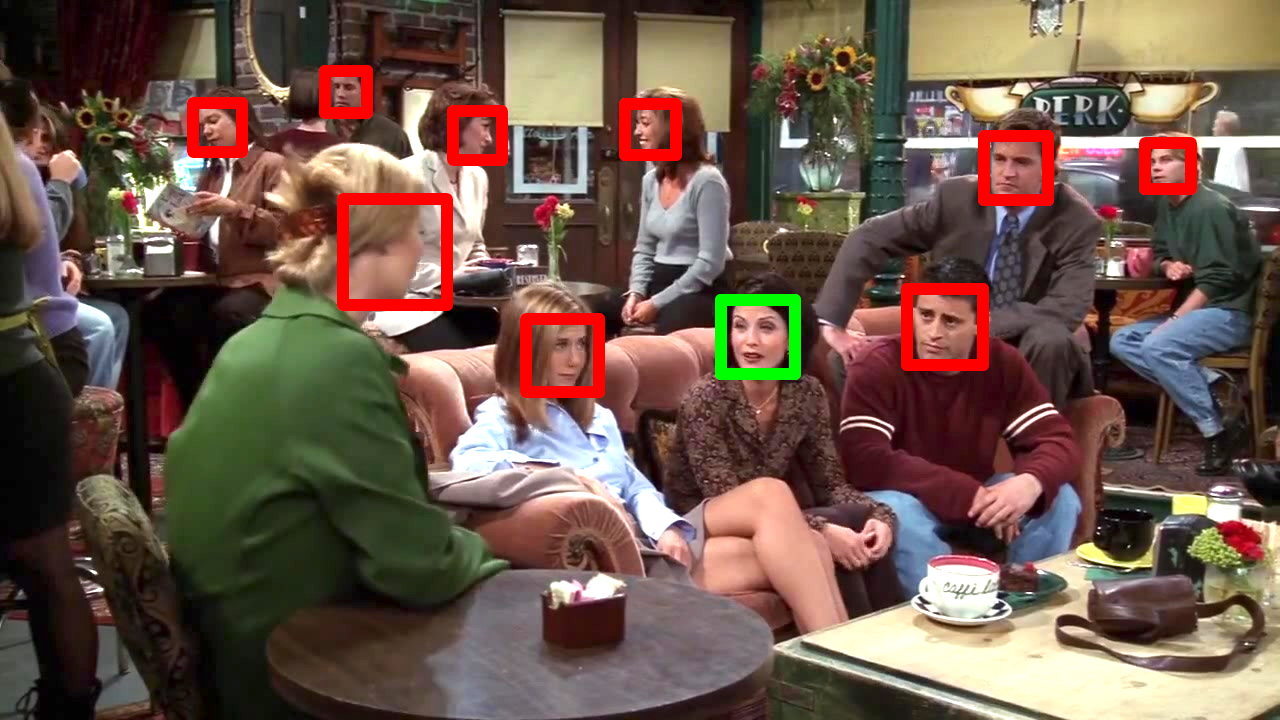}
                }
                \hspace*{\fill}
                \caption{Examples of conversational scenes from MELD videos. Green boxes identify those who are speaking, whereas red boxes mark silent people.}
                \label{fig:asd_meld_examples}
            \end{figure}

            \citet{tao2021} offer a practical implementation of TalkNet-ASD\footnote{\url{https://github.com/TaoRuijie/TalkNet-ASD/blob/main/demoTalkNet.py}}, which we apply to the facial expression and emotional speech data extracted from the realigned MELD videos. In that implementation, each of the face tracks of a given person and the corresponding audio frame sequence are split into blocks and sent to TalkNet-ASD to determine in which frames that given person is actively speaking. Each of those blocks corresponds to a video sequence of up to $\phi$ video frames. Several values for $\phi$ are used in the implementation, namely 25, 50, 75, 100, 125, and 150, as a means to guarantee a more reliable result. A given value of $\phi$ implies that $\phi$ face images and $4 \, \phi$ audio frames in each block are used as input to the TalkNet-ASD model. TalkNet-ASD yields $\phi$ scores $s_{i, j, \phi}$ per block, indicating whether a given person $\pi_{j}$ is detected as actively speaking in frame $f_{i}$ in that block composed of $\phi$ video frames. After getting all scores for every frame, with all different possible values of $\phi$, a resulting score $s_{i, j}$ is obtained by averaging the scores $s_{i, j, \phi}$. A score $s_{i, j} > 0$ indicates that person $\pi_{j}$ is predicted as actively speaking in frame $f_{i}$. Figure~\ref{fig:asd_meld_examples} provides two examples of the application of TalkNet-ASD to the videos of MELD. In both examples, the uttering speakers are marked with green boxes around their faces.
            
            After TalkNet-ASD has generated scores for each face track, the scores are grouped based on their respective tracks to determine which faces belong to the same person. However, if two face tracks have faces from the same frame and both tracks have detected speaking activity, this can result in a ``false positive'', where one of the tracks belongs to someone who is not actively speaking. These face tracks that provide conflicting information on the active speaker are called conflicting face tracks. To reduce false positives, the face tracks are grouped based on the camera cut where they appear. Each group contains a set of face tracks where each track has a conflicting track within the same set. A heuristic is used to eliminate conflicting face tracks according to three criteria:
            \begin{inparaenum}[\itshape i\upshape)]
                \item reduce the number of conflicting face tracks to zero;
                \item maximise the total number of faces associated with speaking activity for all non-conflicting face tracks within a set; and
                \item minimise the number of face tracks, provided the first two criteria are met.
            \end{inparaenum}
            The last criterion is due to the low likelihood of a single person having their extracted face sequence appearing in several face tracks within the same set. After eliminating the conflicting tracks, the remaining non-conflicting tracks are grouped together and ordered based on their associated frame number. The procedure then outputs the resulting sequence of faces, which is associated with the active speaker.
            
            The utterance speaker localisation procedure is executed for every realigned MELD video, which, in turn, is assigned to one particular utterance in the dataset each. Most time consumption derives from the frame-wise face extraction and from the detection of speaking activity in every sequence of facial expressions previously extracted and organised.

    \section{Assessment of the MELD-FAIR Dataset}
    \label{sec:dataset_assessment}
        
        To assess the applicability of MELD-FAIR in ERC, it is important to determine whether the distribution of its data after the dataset refinement procedure is kept similar to that of the original dataset. Two criteria can be used to evaluate whether the data distribution was kept similar to its original distribution. Specific steps of the dataset refinement depend on the target uttering speaker, thus it is desirable that the proportion of utterances in MELD-FAIR assigned to a given speaker remains close to its original proportion in MELD. Similarly, the proportion of utterances assigned to a given emotion should also be kept close to its original proportion, to not alter the task. Moreover, because MELD was built for emotion recognition in conversational contexts, it is worthwhile to determine the portion of dialogues in which the data of at least one utterance was removed during the dataset refinement process.
        
        After assessing whether most of the original utterances are kept in MELD-FAIR, and whether its data distribution is nearly unaltered, it is worthwhile analysing whether the video realignment produces refined speech signals that actually correspond to the speakers provided by the dataset. The retention of many original utterances and the proper correspondence between the speech signals and the expected speakers are indications that the acoustic data is reliable and therefore useful for an application in ERC. Finally, to determine the reliability of the process of localising the uttering speaker, we propose using an emotion recognition model trained on MELD-FAIR and comparing its performance to existing ERC approaches trained on the original version of MELD that use information from visual and/or acoustic modalities. A superior performance of our emotion recognition model would indicate that the emotional facial expressions extracted by the uttering speaker localisation procedure are indeed useful for emotion recognition applications.
        
        \subsection{Properties of the MELD-FAIR Dataset}
        \label{sec:refined_meld_dataset_properties}
            
            The process of dataset refinement consists of two steps, video realignment and utterance source localisation. These refining steps may eventually lead to some utterances of the original MELD dataset not having corresponding audiovisual data in MELD-FAIR. This may happen due to two main reasons. First, the video realignment step may produce an empty video for a given utterance in case the CTC segmentation algorithm determines that in the most likely alignment, $u_{i}$ is aligned to a very small slice of the dialogue audio. Second, even when new video cuts are produced in the video realignment step, no uttering speaker may be located in the scene. Tables~\ref{tab:meld_dataset_refined_emosent_distribution} and~\ref{tab:meld_dataset_refined_speaker_distribution} present the number of dataset records for which there are corresponding audiovisual data in MELD-FAIR, alongside the number of dataset records in its original version. Table~\ref{tab:meld_dataset_refined_emosent_distribution} presents the dataset record distribution according to the annotated emotion and dataset split, and Table~\ref{tab:meld_dataset_refined_speaker_distribution} presents the dataset record distribution according to the utterance speaker and dataset split. Tables~\ref{tab:meld_dataset_refined_emosent_distribution} and~\ref{tab:meld_dataset_refined_speaker_distribution} show that the \textit{dev} and \textit{test} splits each lost approximately 2.5\% of their records in the dataset refinement process. Regarding the \textit{train} split, data loss due to the dataset refinement was also relatively small, with the audiovisual data of MELD-FAIR corresponding to 96.7\% of the utterances of the original MELD dataset.
            
            \begin{table}[ht!]
                \centering
                \caption{Distribution of emotion annotations in the MELD-FAIR dataset. The numbers of original dataset records for each emotion and split are given inside parentheses.}
                \label{tab:meld_dataset_refined_emosent_distribution}
                \begin{tabular}{r|ccc|c}
                    Emotion & \textit{train} & \textit{dev} & \textit{test} & Total \\ \hline
                    neutral & 4537 (4710) & 461 (470) & 1226 (1256) & 6224 (6436) \\
                    joy & 1683 (1743) & 160 (163) & 389 (402) & 2232 (2308) \\
                    surprise & 1158 (1205) & 140 (150) & 270 (281) & 1568 (1636) \\
                    sadness & 670 (683) & 109 (111) & 207 (208) & 986 (1002) \\
                    fear & 261 (268) & 39 (40) & 49 (50) & 349 (358) \\
                    anger & 1082 (1109) & 150 (153) & 339 (345) & 1571 (1607) \\
                    disgust & 267 (271) & 22 (22) & 67 (68) & 356 (361) \\ \hline
                    & 9658 (9989) & 1081 (1109) & 2547 (2610) & 13286 (13708)
                \end{tabular}
            \end{table}
            
            \begin{table}[ht!]
                \centering
                \caption{Distribution of uttering speakers in the MELD-FAIR dataset. The numbers of original dataset records for each speaker and split are given inside parentheses.}
                \label{tab:meld_dataset_refined_speaker_distribution}
                \begin{tabular}{r|ccc|c}
                    Speaker & \textit{train} & \textit{dev} & \textit{test} & Total \\ \hline
                    Rachel & 1392 (1435) & 158 (164) & 350 (356) & 1900 (1955) \\
                    Monica & 1253 (1299) & 130 (137) & 338 (346) & 1721 (1782) \\
                    Phoebe & 1269 (1321) & 183 (185) & 277 (291) & 1729 (1797) \\
                    Joey & 1456 (1509) & 146 (149) & 399 (411) & 2001 (2069) \\
                    Chandler & 1243 (1283) & 100 (101) & 374 (379) & 1717 (1763) \\
                    Ross & 1410 (1459) & 211 (217) & 368 (373) & 1989 (2049) \\
                    others & 1635 (1683) & 153 (156) & 441 (454) & 2229 (2293) \\ \hline
                    & 9658 (9989) & 1081 (1109) & 2547 (2610) & 13286 (13708)
                \end{tabular}
            \end{table}
            
            The data distribution was kept nearly unaltered. For instance, the largest data distribution difference occurred in the fraction of dataset records assigned to the neutral emotion in the \textit{train} split. Out of the original 4710 records in the \textit{train} split that were assigned to the neutral emotion, the dataset refinement procedure was unable to retrieve corresponding audiovisual data for only 173 records. This corresponds to 3.67\% of those records, and to 1.73\% of all records in the \textit{train} split. These dataset records, which correspond to one utterance each, are well dispersed throughout the whole dataset. As a consequence, the fraction of dialogues which lost at least one of their utterances in the dataset refinement procedure is moderately higher. 222 of the 1038 dialogues of the \textit{train} split contain at least one utterance with no corresponding audiovisual data in MELD-FAIR, which represents 21.4\% of the dialogues in that split. For the \textit{dev} and \textit{test} splits, this reduction was lower. 19 of the 114 dialogues of the \textit{dev} split, i.e., 16.7\%, have utterances with no corresponding audiovisual data in MELD-FAIR, and for the \textit{test} split, 49 of its 280 dialogues, i.e., 17.5\%.
        
        \subsection{Assessment of the Video Realignment}
        \label{sec:video_realigment_assessment}
                
                Due to the lack of an annotation of the correct start and end time stamps of each utterance, a self-supervised form of assessing the robustness of the video realignment procedure was devised. A video correctly realigned to its corresponding utterance is expected to have most of its audio content comprised of a speech signal uttered by the speaker annotated in the corresponding dataset record. This would allow training a speaker identification model with the speech signals of the realigned videos of the \textit{train} split so that it generalises and correctly identifies the speakers from speech signals of the realigned videos of the remaining splits. However, the model would require a given speaker to appear in a reasonable number of MELD records in all dataset splits, but only six speakers appear consistently throughout all MELD splits. These are the six main characters: Rachel, Monica, Phoebe, Joey, Chandler, and Ross. The remaining speakers appear rarely, indicating that it is highly unlikely that the speaker identification model could learn to generalise well from their speech.
                
                \subsubsection{Model}
                
                   A speaker identification model is used to assess whether the speech audio in a given 
                   realigned video actually matches the speaker annotated in the corresponding MELD record. The speaker identification model is composed of an encoder part followed by a classifier part. Based on TalkNet-ASD's ATE, a traditional ResNet34 is used as the encoder. This encoder produces an embedding $F_{a}$ of dimensions $\frac{T_{a}}{4} \times 512$, where $T_{a}$ is the number of audio frames corresponding to the speech signal. Then, via temporal max pooling, a 512-dimensional feature vector is obtained for the whole speech signal. Finally, a fully connected layer outputs a prediction regarding the expected speaker of the speech signal from this feature vector.
                
                \subsubsection{Data Augmentation}
                
                    Following the steps of \citet{tao2021}, negative sampling is used to augment the available speech data. In negative sampling augmentation, data is augmented by combining it with some other interfering data within the same batch that effectively shares the same label as the original data, i.e., it is expected that both the original and the interfering speech signal have been uttered by the same speaker. Through randomly selecting interfering data that has those characteristics, an interference is made by combining the original audio tracks and those of the interfering data, thus coming up with a mixture of both. By benefitting from the in-domain noise and the interfering speech signals from the training set itself, this approach presents three advantages in comparison to traditional augmentation through the addition of white noise:
                    \begin{inparaenum}[\itshape i\upshape )]
                        \item the interference data is not artificially generated;
                        \item there is no need for data outside the training set for the audio augmentation; and
                        \item by using audio samples from the same speaker, the interference provided in the data augmentation accentuates the characteristics of that speaker's voice.
                    \end{inparaenum}
                    
                    With a 50\% chance, an audio sample is selected to be augmented this way, which means that within a batch, roughly half of its samples are augmented. Audio samples selected this way are either circularly padded or trimmed to match the size of the original audio sample. A single batch typically has audio samples of very different sizes. In order to let all audio samples in the same batch have the same size, they are either circularly padded or trimmed so that every audio sample in the same batch have a length equal to the average of the lengths of the original audio samples. This way, it is guaranteed that the model is trained with samples of a reasonable size, and that at least half of the samples of a batch consists of unpadded continuous audio samples.
                
                \subsubsection{Training Procedure}
                
                    To train the speaker identification model, audio tracks are randomly sampled, such that there be roughly the same number of audio samples for each class (the six main characters). Audio samples are augmented according to the aforementioned procedure. The model is trained by minimising a cross-entropy loss function using an ADAM optimiser with an initial learning rate of 1e-4, whose value is decreased in half every ten epochs.  Batches of size 64 are used in the model training. The training procedure is kept running until there is a sequence of 30 epochs with no improvement in the weighted F1 score of the \textit{dev} split.
                
                \subsubsection{Results and Analysis}
                    
\begin{figure}[ht!]
                        \centering
                        \hspace*{\fill}
                        	\subfloat[MELD (accuracy: 67.30\%)\label{fig:conf_matrices_speaker_ident_test_original}]{
                        	   \includegraphics[width=0.425\textwidth]{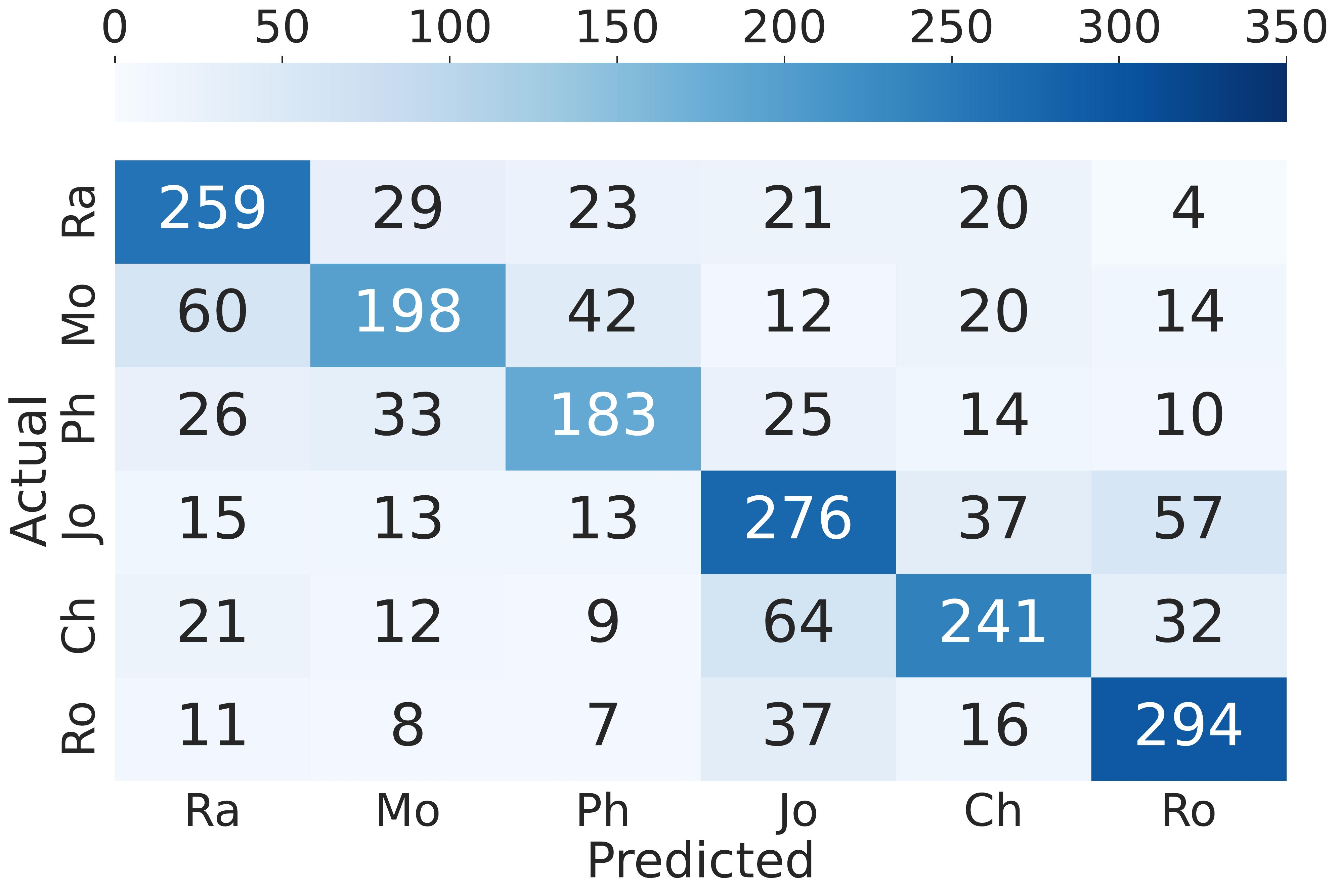}
                        	}\hfill
                        	\subfloat[MELD-FAIR (accuracy: 78.30\%)\label{fig:conf_matrices_speaker_ident_test_refined}]{
                        	   \includegraphics[width=0.425\textwidth]{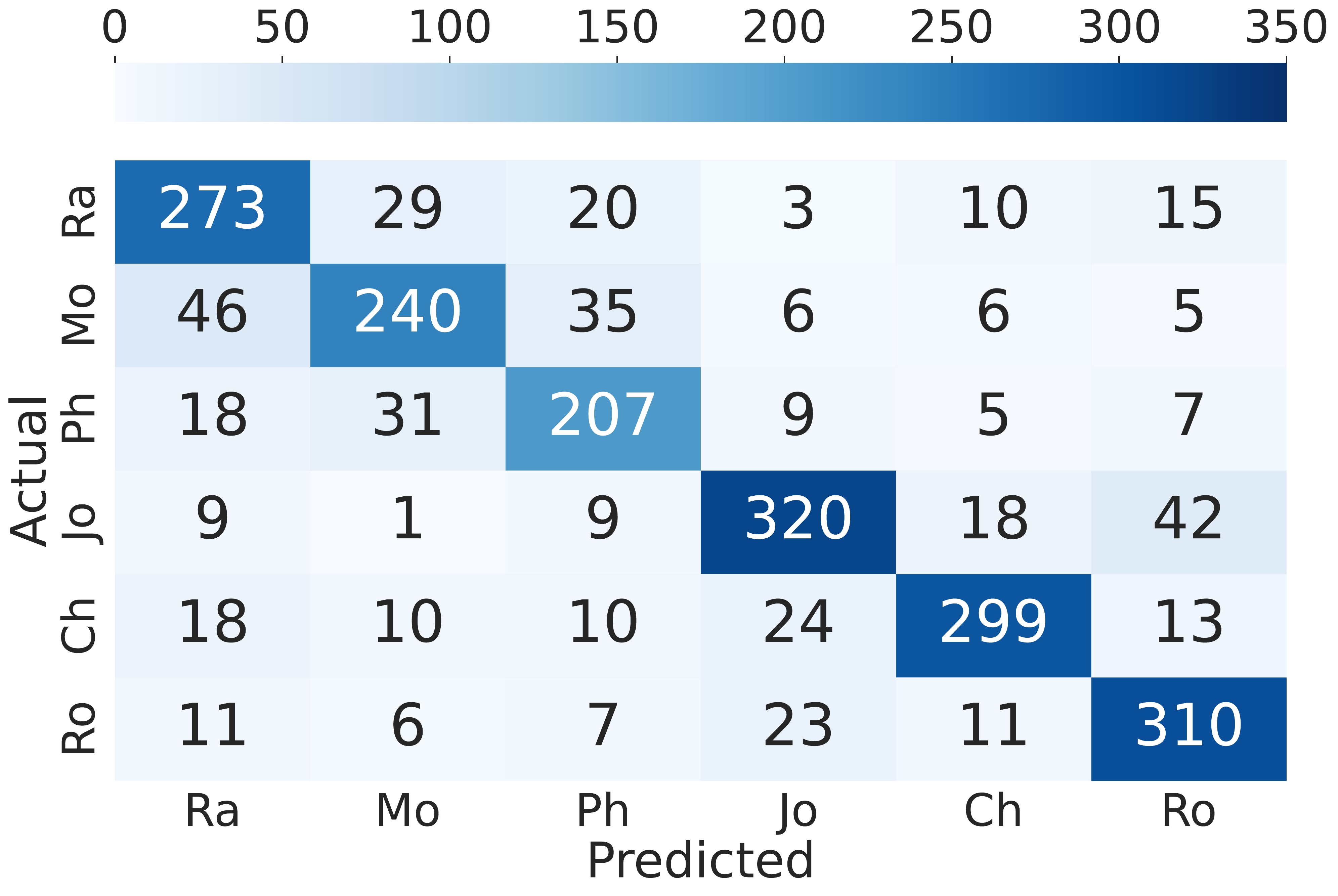}
                        	}
                         \hspace*{\fill}
                        \caption{Confusion matrices of the speaker identification model in MELD's and MELD-FAIR's \textit{test} splits}
                        \label{fig:conf_matrices_speaker_ident}
                    \end{figure}

                    Figure~\ref{fig:conf_matrices_speaker_ident} presents the confusion matrices obtained when evaluating the speaker identification model in MELD's and MELD-FAIR's \textit{test} splits. A comparison is presented on how well the speaker identification model can generalise what it learned from each character's voice from the data of the original MELD dataset (Figure~\ref{fig:conf_matrices_speaker_ident_test_original}) and from that of its refined version, MELD-FAIR (Figure~\ref{fig:conf_matrices_speaker_ident_test_refined}). The speaker identification model subjected to MELD-FAIR achieved a weighted F1 score of 78.32\% in that dataset's \textit{test} split, whereas the speaker identification model subjected to the original MELD achieved a weighted F1 score of 67.07\% in the corresponding \textit{test} split. The confusion matrices and the weighted F1 scores indicate that the video realignment leads to cuts that better match the expected speaker, which, in turn, indicates that it is highly likely that the audio contents of those cuts closely match the corresponding utterances whose transcriptions are given in the dataset.
                    
        \subsection{Application in ERC}
        \label{sec:application_in_erc}
            
            \subsubsection{Model}
                
\begin{figure}[ht!]
                    \centering
                    \includegraphics[width=0.9\textwidth]{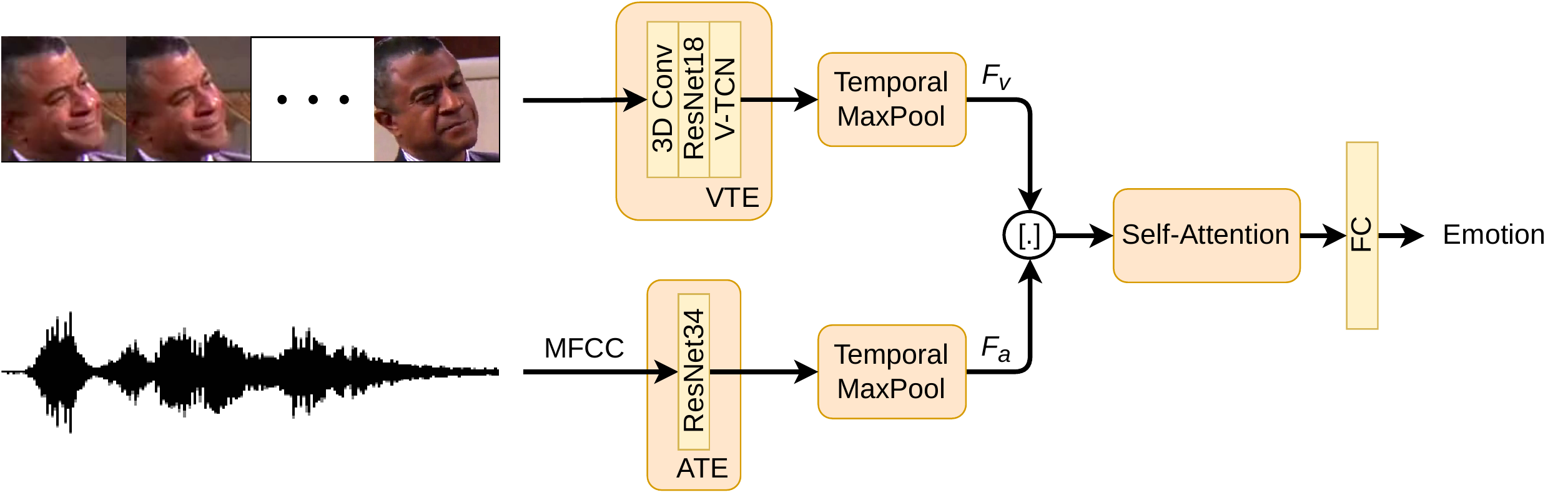}
                    \caption{ERC model}
                    \label{fig:erc_model}
                \end{figure}

                We have devised an emotion recognition (ER) model to assess whether MELD-FAIR actually has visual and acoustic information from which emotional characteristics can be retrieved. Figure~\ref{fig:erc_model} presents the architecture of the ER model. For the encoding of the visual and acoustic inputs, TalkNet-ASD's VTE and ATE have been modified to enable them to produce vector representations with 512 dimensions. {VTE has been modified by having its} sequence of 1D convolution layers removed, since its main application is to reduce the dimensionality of the feature vectors, and V-TCN already yields vector representations with 512 dimensions. For TalkNet-ASD's ATE to produce 512-dimensional feature vectors, its Thin ResNet34 backbone has been changed for a traditional ResNet34. Also, we keep the face crops with their original colour channels for the task of emotion recognition. This way, changes in skin colour due to some emotional reactions, e.g., blushing, can be considered by the ER model. The embeddings output by VTE and ATE are then max-pooled in the temporal dimension into feature vectors $F_{v}$ and $F_{a}$, with 512 dimensions each. These vectors are concatenated and subsequently sent to a self-attention layer. Finally, a fully connected layer yields a prediction for the emotion of the uttering speaker given the output of the self-attention layer.
                
            \subsubsection{Data Augmentation}
                
                Audio samples are augmented through the same data augmentation procedure described in Section~\ref{sec:video_realigment_assessment}. Face crops are augmented by performing one of the following operations: random horizontal flip, random crop of an area with at least 70\% the dimension of the original face crop, or a random rotation up to 15 degrees clockwise or counterclockwise. Afterwards, the face crop is resized to $112 \times 112$ pixels. In order to keep consistency in the direction the speaker's head is looking to, the random characteristics of the data augmentation procedure are applied to the sequence of faces as a whole, and not to each face separately.
                
            \subsubsection{Training Procedure}
                
                Since the distribution of emotion labels is similar in every split of MELD-FAIR, no weighted random sampling in the training of the ER model is performed. Instead, for every record in the \textit{train} split representing a single utterance, a sequence of 15 consecutive face crops is selected as input for the video stream, and the complete utterance audio is provided as input for the audio stream. In case the sequence of faces corresponding to the uttering speaker has less than 15 face crops, then the sequence is circularly padded. If the sequence of faces has more than 15 face crops, then a subsequence of 15 consecutive face crops is randomly selected. The model is trained by minimising a cross-entropy loss function using an ADAM optimiser with an initial learning rate of 1e-4, whose value is decreased in half every ten epochs. Batches of size 64 are used in the model training. The training procedure is kept running until there is a sequence of 30 epochs with no improvement in the weighted F1 score of the \textit{dev} split.

            \subsubsection{Experimental Results}
            
                Three variations of the ER model were implemented and trained from scratch. One incorporated inputs coming from both the acoustic and visual streams, while the other two variations were ablations, each containing only one of the input streams. Table~\ref{tab:er_model_variation_comparison} presents the weighted F1 score achieved by each variation, the number of training epochs it took for every variation to reach its best performance, the average training time per batch, and the number of batches used in each training epoch. The training times presented in Table~\ref{tab:er_model_variation_comparison} were achieved using a single NVIDIA GeForce GTX 1080 Ti.
                
                \begin{table}[ht!]
                    \centering
                    \caption{Comparison of ER model variations}
                    \label{tab:er_model_variation_comparison}
                    \begin{tabular}{r|ccc}
                        Modalities & Vision & Audio & Audio + Vision \\ \hline
                        Weighted F1 score (\%) & 35.58 & 40.54 & 39.81 \\
                        Number of training epochs & 15 & 18 & 19 \\
                        Avg. training time (seconds per batch) & 1.164 & 0.287 & 1.211 \\
                        Number of batches per epoch & 151 & 151 & 151
                    \end{tabular}
                \end{table}
            
            \subsubsection{Comparison with the State of the Art}
            
                To evaluate the benefits of our refinement procedure for the task of ERC with MELD-FAIR, we compare the performance of our ER model to existing approaches that use information from the original MELD videos in ERC, and not only from the utterance transcriptions provided in the dataset.
                
                \textbf{DialogueRNN}~\citep{majumder2019}\footnote{Although DialogueRNN was originally proposed in~\citep{majumder2019}, its first application to ERC in the MELD dataset was in~\citep{poria2019}.} is a baseline approach which models the context of a conversation by tracking the states of individual parties within that conversation. The model determines the emotion of a given utterance according to three aspects: its speaker, the context from preceding utterances, and the emotion thereof. DialogueRNN models these aspects by using three gated recurrent units (GRUs)~\citep{cho2014}, each responsible for a particular aspect.
                
                \textbf{CT+EmbraceNet}~\citep{xie2021} is a pioneering ERC model in using visual information from the MELD videos. Although DialogueRNN predates it, the former uses solely information from the acoustic and textual modalities. This approach uses crossmodal transformers (CTs)~\citep{tsai2019} to enrich the information from one modality by taking into account information from another modality, and this way learn existing correlated information across pairs of modalities. EmbraceNet~\citep{choi2019} was used to carefully deal with the crossmodal information in the feature vectors produced by the crossmodal transformers, and to prevent performance degradation due to the partial absence of data.
                
                \textbf{EmoCaps}~\citep{li2022} uses transformer-based encoders to extract emotion feature vectors from the visual, acoustic and textual modalities. The authors also use BERT~\citep{devlin2019} to extract text feature vectors from every utterance. By concatenating an utterance feature vector with the corresponding emotion vectors of each modality, the authors create a vector representation for that utterance. Then, through the use of a Bi-LSTM~\cite{graves2005,hochreiter1997} and a classification subnetwork, EmoCaps predicts the emotion from every utterance in a dialogue.
                
                \textbf{MMGCN}~\citep{hu2021} uses a multimodal graph, where each node represents a given modality in some particular utterance. Nodes of this graph are connected if they share either the same modality or the same utterance. Each MMGCN node is initialised with a concatenation of two elements: a context-aware feature encoding of the corresponding modality and utterance, and an embedding of the speaker of that particular utterance. MMGCN leverages speaker embedding to inject speaker information into the graph construction. MMGCN encodes the multimodal contextual information through the use of a multilayered deep spectral-domain graph convolutional network. 
                
                \textbf{MM-DFN}~\citep{hu2022}, similarly to MMGCN, uses a multimodal graph with the same structure to characterise the relations between all modalities within a given uttering event, and of every utterance within a dialogue. MM-DFN introduces graph-based dynamic fusion modules, which are stacked in layers, to fuse multimodal context features dynamically and sequentially. These modules aggregate both inter- and intra-modality contextual information in a specific semantic space at each layer. It differs from MMGCN, which aggregates contextual information in a single semantic space. This leads to a gradual accumulation of redundant information. By modelling the contextual information in different semantic spaces, MM-DFN benefits from a reduction in the accumulation of redundant information, as well as from an enhancement in the complementarity between the modalities.
                
                \textbf{M2FNet}~\citep{chudasama2022} is the current state-of-the-art model in ERC in MELD\footnote{Although M2FNet's performance values seem lower than those of other models in Table~\ref{tab:f1_score_comparison}, this is due to most of the contribution in ERC coming from the textual modality, which was not included in Table~\ref{tab:f1_score_comparison}. We decide not to include the performance of those models when the text modality is not ablated because the main objective of this paper is to present a way of extracting useful information from the visual and acoustic modalities, since those are quite unreliable in MELD. In contrast, the text transcriptions are very reliable and do not require an extensive refinement.}. Its main characteristics are
                \begin{inparaenum}[\itshape i\upshape)]
                    \item a visual feature extractor that provides a visual representation based on the faces of the people in a scene as well as on the scene as a whole;
                    \item the use of one stack of transformer encoders for each modality, as a means to learn inter-utterance context on a modality level; and
                    \item a multi-head attention fusion module to better incorporate those modalities, especially the visual and acoustic ones.
                \end{inparaenum}
                
                It is worth noticing that all multimodal approaches to ERC in MELD use context from the dialogue in some form. Since we are interested in extracting the most useful information from the visual and the acoustic modalities, we rely solely on the utterance level. This way, we can guarantee that the performance achieved is a direct consequence of the video realignment and the utterance source localisation, and not from some other part of the dialogue.
                
                Table~\ref{tab:f1_score_comparison} compares the performance of the ER model proposed here with those of ablated versions of all multimodal approaches to ERC in MELD. The values presented in the table were extracted from the literature. Some table cells appear empty because either one modality was not used (e.g., \citet{poria2019} do not use information from the visual modality in their implementation of DialogueRNN), or the authors did not consider the combination of vision and acoustic modalities in their ablation studies (as in~\citep{hu2021} and in~\citep{li2022}). Table~\ref{tab:f1_score_comparison} shows that our ER model achieves a higher weighted F1 score than state-of-the-art approaches when restricted to the visual modality. It is worth noticing that our ER model outperforms state-of-the-art approaches, even though it does not use temporal visual context on a dialogue level. This indicates that the combination of video realignment and active speaker detection can indeed yield sequences of facial expressions which, in turn, provide the ER model with more information on the uttering speaker's emotion than the feature extraction procedures used in the other approaches.
                
                \begin{table}[ht!]
                    \centering
                    \caption{Weighted F1 scores for ERC in MELD \textit{test} split using visual and acoustic data}
                    \label{tab:f1_score_comparison}
                    \begin{tabular}{r|ccc}
                        Model & Vision & Audio & Audio + Vision \\ \hline
                        DialogueRNN & N/A & \textbf{44.3} & N/A \\
                        CT+EmbraceNet & 31.4 & 32.1 & N/A \\
                        EmoCaps & 31.26 & 31.26 & N/A \\
                        MMGCN & 33.27 & 42.63 & N/A \\
                        MM-DFN & 32.34 & 42.72 & \textbf{44.67} \\
                        M2FNet & 32.44 & 39.63 & 35.74 \\ \hline
                        Ours & \textbf{35.58} & 40.54 & 39.81
                    \end{tabular}
                \end{table}
                
                The performance of our ER model when restricted to the acoustic modality is higher than M2FNet (current state-of-the-art approach for ERC in MELD) and EmoCaps. Its performance, however, is lower than those of DialogueRNN, MMGCN and MM-DFN. These models have in common the use of utterance-level feature vectors extracted from OpenSMILE~\citep{eyben2010,schuller2011} as input for the audio stream. EmoCaps also uses these, however, its multimodal representation favours the textual modality since it uses both the utterance feature vector yielded by BERT and an emotion feature vector for the textual modality in its multimodal utterance representation, whereas only a single emotion feature vector is used to represent each of the remaining modalities. Also, EmoCaps's weighted F1 scores in both modalities correspond to that of a model that outputs \textit{neutral} for every input. M2FNet, on the other hand, uses a novel feature extractor module based on the triplet loss~\citep{schroff2015} to fetch deep features from acoustic and visual contents.

    \section{Discussion and Conclusion}
    \label{sec:discussion}
    
        Connectionist-temporal-classification segmentation and active speaker detection allowed us to refine  MELD, a largely-used multimodal dataset for emotion recognition in multi-party conversational scenarios, making it possible to better align its audiovisual data with the corresponding utterance transcriptions, as well as to obtain reliable face crops of the uttering speaker of nearly every scene. The comparison with state-of-the-art approaches also indicates that those face crops provide more precise information on the emotion of the uttering speaker than the most recent approaches.
        
        The reliable extraction of the speakers' face crops from well-realigned videos accounts for the high performance of the vision-only version of our emotion recognition model, which outperforms other competing approaches by more 2.3\%. The relatively simple architecture of our emotion recognition model, as well as its restriction to working on an utterance level, i.e., without contextual information from the whole dialogue, indicate that much of its high performance is due to the improvement in the information from MELD's visual modality.
        
        Furthermore, researchers on emotion recognition in multi-party conversational scenarios can benefit from MELD-FAIR, the refined version of MELD delivered in this publication. More generally, with the recent advancements in deep learning, creating a dataset automatically becomes within sight. Automatic speech recognition allows automatic text transcription, while automatic lip reading, which requires active speaker detection, could verify its correctness, and vice versa.

    \section*{Acknowledgement}
        
        The authors acknowledge partial support from the German Research Foundation DFG under project CML (TRR 169).
    
    \appendix
    
        \section{Problematic cases of MELD}
        \label{sec:app_problematic}
            
            MELD presents a variety of problematic cases beyond the misalignment between the videos and the utterance transcriptions. These comprise multiple other problems which raised errors during the processing of data refinement. Table~\ref{tab:problematic_cases} offers an extensive list of such cases, identified by the split, dialogue id and utterance id of each case.
            
            \begin{table}[!ht]
                \centering
                \caption{List of existing problematic cases in MELD}
                \label{tab:problematic_cases}
                \begin{tabular}{ccc|m{10cm}}
                    Split & Dia ID & Utt ID & \begin{tabular}{c}Problem\end{tabular} \\ \hline
                    \textit{train} & 125 & 3 & \begin{tabular}{l}Corrupted video file\end{tabular} \\ \hline
                    \textit{dev} & 110 & 7 & \begin{tabular}{l}Non-existent video file\end{tabular} \\ \hline
                    \textit{test} & 38 & 4 & \multirow{2}{*}{\begin{tabular}{l}Very long video ($>$ 45 seconds), incompatible with\\[.4\baselineskip]its utterance transcription\end{tabular}} \\
                    \textit{test} & 220 & 0 & \\ \hline
                    \textit{train} & 309 & 0 & \multirow{7}{*}{\begin{tabular}{l}Utterance transcription contains not only the utterance\\[.4\baselineskip]but also a description within parentheses\end{tabular}} \\
                    \textit{train} & 404 & 15 &  \\
                    \textit{train} & 736 & 4 &  \\
                    \textit{train} & 832 & 3 &  \\
                    \textit{train} & 1018 & 2 &  \\
                    \textit{dev} & 108 & 0 &  \\
                    \textit{test} & 128 & 2 &  \\ \hline
                    \textit{train} & 65 & 3 & \multirow{3}{*}{\begin{tabular}{l}Utterance transcription contains not only the utterance\\[.4\baselineskip]but also a description within brackets\end{tabular}} \\
                    \textit{train} & 761 & 1 & \\
                    \textit{test} & 86 & 3 & \\ \hline
                    \textit{train} & 739 & 14 & \multirow{2}{*}{\begin{tabular}{l}No utterance. Just a description within parentheses\end{tabular}} \\
                    \textit{train} & 849 & 3 & \\ \hline
                    \textit{train} & 111 & N/A & \begin{tabular}{l}Utterances not chronologically ordered\end{tabular} \\ \hline
                    \textit{train} & 446 & 19 & \begin{tabular}{l}Should be the first utterance of dialogue 447\end{tabular}
                \end{tabular}
            \end{table}
    
    \bibliographystyle{abbrvnat}
    \bibliography{bibliography}
    
    
    
    
    


\end{document}